\documentclass[epj]{svjour}
\onecolumn
\usepackage{latexsym}
\usepackage{graphics}
\usepackage{amsmath}
\usepackage{amsfonts}
\usepackage{amssymb}
\usepackage{geometry}
\usepackage[english]{babel}
\begin{document}
\title{Quantum Cosmology via Quantization of Point-Like Lagrangian}
\author{Miao He \and Zi-liang Wang \and Jian-bo Deng
\thanks{\emph{Corresponding author:} dengjb@lzu.edu.cn} 
\and Hua Chen% etc
}                     % Do not remove
\institute{Institute of Theoretical Physics, LanZhou University, Lanzhou 730000, P. R. China}
\abstract{
    The purpose of this paper is to introduce a new way to inquire the
    quantum cosmology for a certain gravitational theory. Normally,
    the quantum cosmological model is introduced as the minisuperspace
    theory which is obtained by reducing the superspace where the
    Wheel-DeWitt equation is defined on using the symmetry provided by
    cosmological principle. Unlike that, the key of our approach is to
    reinterpret the cosmology in a classical dynamical way using a
    point-like Lagrangian and then quantize the point-like model. We
    apply the method into Einstein gravity, gravity with a
    cosmological constant and the $f(R)$-gravity, and get their wave
    equations respectively. By analsysing the exact solution for the
    quantum cosmology with and without a cosmological constant we
    demonstrate that the cosmological constant is essential and being a
    tiny positive number. We also show the possibility of explaining
    inflation under the quantum version of cosmology.
 % end of PACS codes
} %end of abstract
\maketitle
\section{Introduction}
\label{intro}
  The motive of studying a quantum cosmological theory mainly emerges
  from two parts. First of all, classical gravity theory fails in
  precisely describing topics like the very early universe. To
  understand those tiny scale objects with huge energy, a successful
  quantum gravity theory is requisite and then we can apply it into
  the cosmological case. However since no such theory is available
  till now, one considerable effort is to view the quantum cosmology
  as an effective theory which can be approximately obtained by
  modifying the classical theory.

  In classical field theory, the Lagrangian density is a function of one or
  more
  fields and their derivatives. Then one gets the action
    \begin{equation}
    \label{eq:action}
    S=\int L\,dt=\int\mathcal{L}(\phi,\partial_\mu\phi)\,d^4x \qquad.
  	 \end{equation}
  According to the principle of least action, one
  can get corresponding equations like Maxwell equations, Klein-Gordon
  equation, Dirac equation etc. Similarly, in the
  general relativity, the Einstein-Hillbert actions discribed as
  	 \begin{equation}
    \label{eq:E-H action}
    S_{EH}=\int\mathcal{L}\,d^4x=\int \sqrt{-g}R\,d^4x \qquad.
  	 \end{equation}
  which can get the Einstein field equation. This Lagrangian formulation
  leads to get
  Hamiltionian formulation, through Legendre transformation, which would make
  the transition to quantum mechanics easier.
  An attempt following this idea is called the
  minisuperspace quantization which is first purposed by
  DeWitt~\cite{dewitt_quantum_1967}.
  At that time Wheeler had
  introduced the idea of superspace, the space of all three-geometries
  as the arena in which geometrodynamics develops, a
  particular four-geometries being represented by a trajectory in this
  space. Misner had just finished applying the Hamiltonian formulation
  of gravity, developed in the late 1950s and early 1960s, to
  cosmological models with an eye towards quantization of these
  cosmologies as model theories of general
  relativity~\cite{misner_quantum_1969}, which is another motive for
  studying quantum cosmology. He invented ``minisuperspace'' and
  ``minisuperspace quantization'' or ``quantum cosmology'', to describe
  the evolution of cosmological spacetimes as trajectories in the
  finite-dimensional sector of superspace related to the finite number
  of parameters, which describe $t=$const slices of the models, and the
  quantum version of such models, respectively.

  For the classical cosmology, its dynamic equation is the Friedman equation 
  which is derived from Einstein field equation. Under standard model, 
  for the quantum cosmology, a quantum gravity theory might be an effective approach.
  Cosmological minisuperspaces and their quantum versions were
  extensively studied in the early 1970s, but interest in them waned
  after about 1975 and little new work was done until Hawking revived
  the field in the 1980s~\cite{hawking_boundary_1982,hartle_wave_1983},
  emphasizing path-integral approaches. This
  started a lively resurgence of interest in minisuperspace
  quantization till now. For a quick review of the minisuperspace
  quantization see Halliwell~\cite{halliwell1,halliwell2,halliwell3}.

  In this paper, we would like to introduce a new technique in
  acquiring the quantum system of cosmology via the point-like
  Lagrangian under a certain gravitational model. According the cosmological
  principle, if we use the spacial symmetry of our univese first,
  a point-like Langrangian can be got as bellow
  	 \begin{equation}
     \label{eq:point-like lag}
      L= \int \mathcal{L}\,d\Omega=\int \sqrt{-g}R\,d\Omega \qquad,
  	 \end{equation}
  where $g$ is the determinant of Friedmann-Robertson-Walker(FRW) metric, 
  $R$ is the scalar curvature and $d\Omega$ 
  represents the spacial volume element.
  The Ensitein-Hillbert action can be written as
    \begin{equation}
    \label{eq:point-like lag}
      S = \int L\,dt \qquad,
  	 \end{equation}
  here $t$ is the cosmological time,
  which can be treated as a classical mechanics action.
  Then quantize this model through the classical  	
  canonical quantization. It is
  much easier than the normal process but reaches the similar result. We
  will then apply this technique into a more complicated case, which
  contains a cosmological constant, and even more general, for the $f(R)$
  gravity. We will give the exact solution to the quantum universe
  with a cosmological constant and discuss its meaning.

\section{The Friedmann Equations}
  In the normal sense, we get the cosmological model under a certain
  kind of gravity by applying the cosmological principle into the
  gravitational field equation which is obtained from variation of the
  action. That is to say, considering a action with matter field
  \begin{equation}
   \label{eq:action_matter}
   S=\int\sqrt{-g}R\,d^4x+S_M \qquad,
  \end{equation}
  by varying the action eq.\eqref{eq:action_matter} with respect to $g_{\mu\nu}$, one can get 
  the Enistein Field Equation. After setting the metric in the equation to the FRW metric 
    \begin{equation}
    \label{eq:frw-metric}
    ds^2 = -dt^2 +a^2(t) \left(\frac{dr^2}{1-\kappa r^2}
      +r^2(d\theta^2+\sin^2\theta d\varphi^2)\right) \qquad,
  \end{equation}
  and the energy-momentum tensor with a perfect fluid
  \begin{equation}
  T^{\mu\nu}=(-g)^{-1/2}\frac{\delta S_M}{\delta g_{\mu\nu}}=(p+\rho)U^{\mu} U^{\nu}+pg^{\mu\nu}\qquad, 
  \end{equation}
  it 
  would end up with the Friedman Equations.
  
  Surprisingly, one can apply the cosmology principle directly into the action before variation.
  This modifies the Lagrangian to a classical point-like one with respect to
  the scale factor. By the variation of the point-like action, it will give exactly the
  same equations as those in the normal process. Such method has been
  extensively used in studying scalar field cosmology~\cite
  {PhysRevD.42.1091,PhysRevD.69.103510,PhysRevD.83.103512}, non-minimally
  coupled cosmology~\cite{PhysRevD.62.043506,sanyal_general_2003},
  scalar-tensor theory~\cite{PhysRevD.52.3288}, multiple scalar
  fields~\cite{10.4137/GBI.S4273}, vector
  field~\cite{0264-9381-27-13-135019}, Fermion
  field~\cite{0264-9381-25-22-225006}, $f(T)$
  gravity~\cite{wei_noether_2012}, $f(R)$
  gravity~\cite{capozziello_fr_2008}, high order gravity
  theory~\cite{capozziello_general_2000}, Gauss–Bonnet
  gravity~\cite{sanyal_general_2011} and so on.
  Without loss of generality, we get the action as below by setting $\kappa=0$, 
  \begin{equation}
   \label{eq:action_matter-2}
   S=\int-6a\dot{a}^2dt +S_M \qquad,
  \end{equation}
  the variation with respect to $a$ and take note of 
  $\delta S_M/\delta a=\delta S_M/\delta g_{\mu\nu}\cdot\delta g_{\mu\nu}/\delta a=6p/a$, one    
  can get one of the fridemann equation
   \begin{equation}
   \label{eq:friedmann_equation_1}
   2\frac{\ddot{a}}{a}+\frac{\dot{a}^2}{a}=-p \qquad.
  \end{equation} 
  By combining the conservation equation of perfect fluid $\dot{\rho}=-3(p+\rho)\dot{a}/a$, 
  all Friedmann equations can be derived from this point-like Lagrangian. 
  
  Moreover, this point-like Lagrangian may lead us to quantiza the cosmology, one would get an 
  easier wave equation and its analytic solutions. In addition, through the classical
  canonical quantization, the evolution of the cosmology which can be discribed by the scale 
  factor, may be different from the classical cosmology as it has the quantum effects. Next, we 
  will apply this method into the quantum cosmology.
  
\section{Point-Like Quantum Cosmology}
  For simplicity, let us restrict our discussion to the flat FRW
  universe($\kappa=0$) in the case of without considering the matter field. 
  Recall the point-like Lagrangian for a 
  flat FRW universe under Einstein gravity, its effective 
  part is
  \begin{equation}
    \label{eq:point-R-lagrangian}
    L = -6a\dot{a}^2 \qquad.
  \end{equation}
  Its canonical momentum is
  \begin{equation}
    \label{eq:point-R-momentum-a}
    \pi_a = \frac{\partial L}{\partial \dot{a}} =
    -12a\dot{a} \qquad,
  \end{equation}
  from which we can get its Hamiltonian
  \begin{equation}
    \label{eq:point-R-hamiltonian}
    H = \pi_a\dot{a} -L = -6a\dot{a}^2 =
    -\frac{\pi_a^2}{24a} \qquad.
  \end{equation}
  Following the process of standard canonical quantization,
  according to the Weyl rule~
  \cite{weyl1950theory}, we can
  get the Hamiltonian operator
  \begin{equation}
    \label{eq:point-R-hamiltonian operator}
    \hat{H}=-\frac{1}{24}\frac{1}{4}(\hat{\pi}_{a}^2\frac{1}{a}	
    +2\hat{\pi}_{a}\frac{1}{a}\hat{\pi}_{a}+\frac{1}{a}\hat{\pi}_{a}^2)
    \qquad,
  \end{equation}
  then we should replace the canonical momentum $\pi_a$ by the operator of
  momentum
  $-i\partial_a$ and get the wave equation of the scale factor $a$
  \begin{equation}
    \label{eq:point-R-schrodinger}
    i\frac{\partial\psi}{\partial t}=\frac{1}{24}(\frac{1}{a}\frac{\partial^2
    \psi}{\partial{a^2}}-\frac{1}{a^2}\frac{\partial\psi}{\partial{a}}+
    \frac{1}{2a^3}\psi) \qquad.
  \end{equation}
  Since the scale factor $a$ is a non-negative real number in
  cosmology, this wavefunction should be a function defined only on
  the right half of the real line with its value being a complex
  number at any given time.

  Assuming $\psi(a,t)$ takes the form of
  $e^{-i\epsilon t}\phi(\epsilon;a)$ and pluging it into 
  eq.\eqref{eq:point-R-schrodinger}, we get the eigen equation
  \begin{equation}
    \label{eq:point-R-energy}
    \frac{1}{24}[\frac{1}{a}\phi''(a)-\frac{1}{a^2}\phi'(a)+\frac{1}{2a^3}
    \phi(a)]=\epsilon\phi(a) \quad,\quad a\geqslant
    0 \qquad.
  \end{equation}
  The solution of this equation depends on whether the eigenvalue is
  positive or not.

  When $\epsilon>0$ we can first rescale the variable $\xi=(24\epsilon)^{1/3}
  a$ to drop the
  parameters in eq.\eqref{eq:point-R-energy} and modify it to
  \begin{equation}
    \label{eq:point-R-energy-modified-pos}
    \xi^2\phi''-\xi\phi'+(\frac{1}{2}-\xi^3)\phi=0 \quad,\quad \xi\geqslant 0
    \qquad.
  \end{equation}
  Through a variable substitution of $z=2\xi^{3/2}/3$ and introducing a new
  function $u(z)$ with the relationship
  $\phi(z)=\xi\,u(z)$, we get
  \begin{equation}
    \label{eq:modified-bessel}
    u''(z)+\frac{1}{z}u'(z)-(1+\frac{(\sqrt{2}/3)^2}{z^2})u(z)=0 \qquad,
  \end{equation}
  which is a modified Bessel equation whose solutions can be
  described by a linear combination of the first modified Bessel
  function $I_{\sqrt{2}/3}(z)$ and the second modified Bessel function
  $K_{\sqrt{2}/3}(z)$.

  However $I_{\sqrt{2}/3}(z)$ increases in exponential form when $z$
  goes to
  infinity, which shows great divergent trend and can not be accepted
  as a reasonable wavefunction. Therefore, recovering all the
  transformations we made, we get the eigen function with a given
  eigenvalue described by eq.\eqref{eq:point-R-energy} of
  \begin{equation}
    \label{eq:point-R-wave-pos}
    \phi^+(\epsilon;a)=(24\epsilon)^{\frac{1}{3}}aK_{\sqrt2/3}(\frac{4\sqrt{6
    \epsilon}}{3}a^{\frac{3}{2}}) \qquad.
  \end{equation}

  The asymptotic expansion of the second modified Bessel function with
  huge arguments can be described as
  \begin{equation}
    \label{eq:modified-bessel-2-asymp-huge}
    K_\nu(z) \sim \sqrt{\frac{\pi}{2z}}\,e^{-z} \qquad,
  \end{equation}
  which indicates our wavefunction $\phi^+$ behaves like
  \begin{equation}
    \label{eq:point-R-wave-pos-asymp-huge}
    \phi^+(\epsilon;a)\sim a^{\frac{1}{4}}e^{-4\sqrt{6\epsilon}a^{3/2}/3}
    \qquad,
  \end{equation}
  and will decay to zero at an extremely quick rate with $a$ getting
  huge enough. This means $\phi^+$ is normalizable and is suitable for
  being a wavefunction.

  For tiny arguments, the second modified Bessel function has the
  following asymptotic form,
  \begin{equation}
    \label{eq:modified-bessel-2-asymp-tiny}
    K_\nu(z) \sim \frac{\Gamma(\nu)}{2} \left(\frac{2}{z}\right)^\nu
    \qquad.
  \end{equation}
  Applying this into the wavefunction eq.\eqref{eq:point-R-wave-pos},
  one can find when $a\rightarrow 0$, the wavefunction
  \begin{equation}
    \label{eq:point-R-wave-pos-asymp-tiny}
    \phi^+(\epsilon;a)\sim a^{(2-\sqrt2)/2}
  \end{equation}
  goes zero which can avoid the possibility of a
  cosmological singularity.

  When $\epsilon<0$, in order to keep $\xi$ being positive, we should
  rescale the variable in form of $\xi=(-24\epsilon)^{1/3}a$. Now
  equation eq.\eqref{eq:point-R-energy} becomes
  \begin{equation}
    \label{eq:point-R-energy-modified-neg}
    \xi^2\phi''-\xi\phi'+(\frac{1}{2}+\xi^3)\phi=0 \quad,\quad \xi\geqslant 0
    \qquad.
  \end{equation}
  Again we apply the substitution $z=2\xi^{3/2}/3$ and let
  $\phi=\xi\,u$, then we end up with a Bessel equation
  \begin{equation}
    \label{eq:bessel}
    u''(z)+\frac{1}{z}u'(z)+(1-\frac{(\sqrt{2}/3)^2}{z^2})u(z)=0 \qquad.
  \end{equation}
  Normally we choose the Bessel function of first kind $J_{\sqrt{2}/3}(z)$ as
  one of the basis for its solution space. The other base can have
  different choice among which the most convenient should be the
  Bessel function of second kind $Y_{\sqrt{2}/3}(z)$ also known as the Weber
  function or the Neumann function.

  So for every distinct eigenvalue, the quantum system has the
  degenerate degree of two with its independence basis chosen as
  \begin{eqnarray}
    \label{eq:point-R-wave-neg-1}
    \phi^-_1(\epsilon\,;a) &= &(-24\epsilon)^{\frac{1}{3}}a\,
    J_{\frac{\sqrt2}{3}}(\frac{4\sqrt{-6\epsilon}}{3}\,a^{\frac{3}{2}})
    \qquad,\\
    \label{eq:point-R-wave-neg-2}
    \phi^-_2(\epsilon\,;a)&= &(-24\epsilon)^{\frac{1}{3}}a\,
    Y_{\frac{\sqrt2}{3}}(\frac{4\sqrt{-6\epsilon}}{3}\,a^{\frac{3}{2}})
    \qquad.
  \end{eqnarray}

  For great arguments, the Bessel functions behave like the following
  \begin{eqnarray}
    \label{eq:bessel-1-asymp-huge}
    J_\nu(z) &\sim &\sqrt{\frac{2}{\pi z}} \left(\cos\left(z
                     -\frac{\nu\pi}{2} -\frac{\pi}{4}\right)
                     +O(z^{-1})\right) \qquad, \\
    \label{eq:bessel-2-asymp-huge}
    Y_\nu(z) &\sim &\sqrt{\frac{2}{\pi z}} \left(\sin\left(z
                     -\frac{\nu\pi}{2} -\frac{\pi}{4}\right)
                     +O(z^{-1})\right) \qquad.
  \end{eqnarray}
  Therefore, when $a$ goes to infinity, the wavefunctions have
  asymptotic expansions
  \begin{eqnarray}
    \label{eq:point-R-wave-neg-1-asymp-huge}
    \phi^-_1(\epsilon\,;a) &\sim &a^{\frac{1}{4}}
                                   \cos\left(\frac{4\sqrt{-6\epsilon}}{3}\,
                                   a^{\frac{3}{2}}
                                   -\frac{2\sqrt2+3}{12}\pi\right)
                                   +O(a^{-\frac{5}{4}}) \qquad,\\
    \label{eq:point-R-wave-neg-2-asymp-huge}
    \phi^-_2(\epsilon\,;a) &\sim &a^{\frac{1}{4}}
                                   \sin\left(\frac{4\sqrt{-6\epsilon}}{3}\,
                                   a^{\frac{3}{2}}
                                   -\frac{2\sqrt2+3}{12}\pi\right)
                                   +O(a^{-\frac{5}{4}}) \qquad,
  \end{eqnarray}
  that somehow behaves like sine-cosine with a raise to the power of
  $1/4$. thus $\phi^-_1,\phi^-_2$ are divergent when $z$ goes to infinity,
  which cannot be accepted as a reasonable wavefunction.

  As shown from our discussion, only $\phi^+$ can act as the wavefuntion and
  is capable of avoiding the cosmological singularity. Its eigenvalue $
  \epsilon$
  would be a continue spectrum which can has value from zero to
  infinity. Normally, for any initial states $\sum_{\epsilon}c_{\epsilon}|
  \epsilon\rangle$(here $|\epsilon\rangle$ is the nomalized eigen state). the
  expectation value of $a^2$ evolves like
	 \begin{equation}
		 \label{eq:scale-factor-evolve}
			\overline{a^2}(t)=\sum_{\epsilon\neq\epsilon'}c_\epsilon c_{\epsilon'}
			\langle\epsilon|a^2|\epsilon'\rangle cos((\epsilon-\epsilon')t+\theta_
			{\epsilon,\epsilon'})+\sum_
			{\epsilon}|c_\epsilon|^2\langle\epsilon|a^2|\epsilon\rangle \qquad.
	 \end{equation}
	Therefore, the evolution of $a^2$ can be any functions with specific initial
	state, including the inflation of the early universe. However, many theory
	consider the cosmology constant is essential, like $\Lambda$CDM model and
	Quantum Field Theory. So we would like to consider a quantum cosmological model
	under some more generalized gravitational models like gravity with a cosmological constant.
	
\section{Quantum Cosmology with A Cosmological Constant}
\label{}
The point-like Lagrangian for flat FRW universe under the
  gravitational model with a cosmological constant has the form
  \begin{equation}
    \label{eq:point-lambda-lagrangian}
    L = -6a\dot{a}^2 +\Lambda a^3 \qquad.
  \end{equation}
  The additional term of $\Lambda a^3$ does not contain the derivative
  of $a$, thus will not change the form of canonical momentum we get in
  eq.\eqref{eq:point-R-momentum-a}. So the Hamiltonian in this case is
  \begin{equation}
    \label{eq:point-lambda-hamiltonian}
    H = -\frac{\pi_a^2}{24a} -\Lambda a^3 \qquad.
  \end{equation}
  The same quantization method we can get the wave equation
  \begin{equation}
    \label{eq:point-lambda-schrodinger}
    i\frac{\partial\psi}{\partial t}=\frac{1}{24}(\frac{1}{a}\frac
    {\partial^2\psi}{\partial{a^2}}-\frac{1}{a^2}\frac{\partial\psi}
    {\partial{a}}+\frac{1}{2a^3}\psi)-\Lambda a^3\psi \qquad.
  \end{equation}
  Its eigen equation is given by
  \begin{equation}
    \label{eq:point-lambda-energy}
    \phi''-\frac{1}{a}\phi'+(\frac{1}{2a^2}-24\Lambda a^4-24\epsilon
    a)\phi=0\qquad.
  \end{equation}
  Next we will consider the asymptotic behavior of the equation.

  When $a\rightarrow 0$, the wave equation can be described as
	\begin{equation}
		\label{eq:point-lambda-energy-zero}
		\phi''-\frac{1}{a}\phi'+\frac{1}{2a^2}\phi=0\qquad,
	\end{equation}
	obviously,
	\begin{equation}
		\label{eq:point-lambda-wave-zero}
		\phi\sim a^{1+\sqrt{2}/2},a^{1-\sqrt{2}/2}\qquad.
	\end{equation}
	
	When $a\rightarrow\infty$, the equation behaves like
	\begin{equation}
		\label{eq:point-lambda-energy-infinity}
		\phi''-(24\Lambda a^3+24\epsilon a)\phi=0\qquad,
	\end{equation}
  it has a special solution with the form like
  	\begin{equation}
		\label{eq:point-lambda-wave-infinity}
		\phi\sim e^{\nu a^3}\qquad.
	\end{equation}

  Generally, assuming $\phi(a)$ takes the forms like 
	 \begin{equation}
	 		\label{eq:point-lambda-wave-form-1}
			\phi(a)=a^{1+\sqrt{2}/2}\cdot e^{\nu a^3}\cdot u(a)\qquad,
		\end{equation}
	or
		\begin{equation}
			\label{eq:point-lambda-wave-form-2}
			\phi(a)=a^{1-\sqrt{2}/2}\cdot e^{\nu a^3}\cdot u(a)\qquad.
		\end{equation}
	 Applying eq.\eqref{eq:point-lambda-wave-form-1} into the eigen
	 equation eq.\eqref{eq:point-lambda-energy} and setting $9\nu^2=24
	 \Lambda$, one would get the equation for $u(a)$ as bellow
	\begin{equation}
		\label{eq:point-lambda-energy-modified}
	au''(a)+(6\nu a^{3}+1+\sqrt{2})u'(a)+[(9+3\sqrt{2})\nu-24		
	\epsilon]a^2u(a)=0\qquad.
	\end{equation}
	With substitution $z=-2\nu a^3$, rearrange
	eq.\eqref{eq:point-lambda-energy-modified}
	\begin{equation}
	\label{eq:kummer-equation}
	zu''(z)+(\frac{3+\sqrt{2}}{3}-z)u'(z)-(\frac{3+\sqrt{2}}{6}-\frac{4	
	\epsilon}{3\nu})u(z)=0\qquad,
	\end{equation}
  which is a confluent hypergeometric equation that is also
  known as Kummer's equation. It has a solution described by Kummer's
  function defined as
  \begin{equation}
    \label{eq:kummer-function}
    F(\alpha,\gamma,z) = \sum_{n=0}^\infty
    \frac{\alpha^{(n)}}{\gamma^{(n)}} \frac{z^n}{n!} \qquad,
  \end{equation}
  where in our case $\alpha=(3+\sqrt2)/6-4\epsilon/3\nu$ and
  $\gamma=(3+\sqrt2)/3$. Here
  the symbol $x^{(n)}$($x=\alpha,\gamma$) stands for a rising factorial defined as
  \begin{eqnarray}
    x^{(0)} &= &1 \qquad,\\
    x^{(n)} &= &x(x+1) \cdots (x+n-1) ,\quad n\geqslant 1 \qquad.
  \end{eqnarray}
  Since $\gamma$ is not a integer, another solution independent with 
  eq.\eqref{eq:kummer-function} can be simply introduced as
  \begin{equation}
    u(z) = z^{1-\gamma} F(\alpha-\gamma+1,2-\gamma,z) \qquad.
  \end{equation}

  Recovering from the substitution we made and considering $\nu$ can be
  either positive or negative, it seems that we will get four independent
  eigen functions for any given eigenvalue $\epsilon$
  \begin{eqnarray}
    \label{eq:point-lambda-wave-1}
    \phi^{\pm}_1(\epsilon\,;a) &= &a^{1+\sqrt2/2}e^{\pm\nu a^3}
    																				F(\frac{3+\sqrt2}{6}\mp \frac{4\epsilon}
    																				{3\nu},\frac{3+\sqrt2}{3},\mp2\nu a^3)
                                  \qquad,\\
    \label{eq:point-lambda-wave-2}
    \phi^{\pm}_2(\epsilon\,;a) &= &a^{1-\sqrt2/2}e^{\pm\nu a^3}F(\frac{3-
    																				\sqrt2}{6}\mp \frac{4\epsilon}{3\nu},
    																				\frac{3-\sqrt2}{3},\mp2\nu a^3)
                                 \qquad,
  \end{eqnarray}
  where $\nu=2\sqrt{6\Lambda}/3$. However Kummer's function obeys the
  property of Kummer's transformation
  \begin{equation}
    \label{eq:kummer-transformation}
    F(\alpha,\gamma,z) = e^zF(\gamma-\alpha,\gamma,-z) \qquad.
  \end{equation}
  So in fact one can check that $\phi^+_{1,2}=\phi^-_{1,2}$, and there
  are only two independent solutions for each $\epsilon$. In addtion,
  when apply eq.\eqref{eq:point-lambda-wave-form-2} into
  eq.\eqref{eq:point-lambda-energy} we get the same two solutions
  except a constant coefficient according to Kummer transfermation.

  For great arguments, Kummer's function can be approximately expanded
  as
  \begin{equation}
    \label{eq:kummer-equation-asymp-huge}
    F(\alpha,\gamma,z) \sim \Gamma(\gamma)
    \left(\frac{e^zz^{\alpha-\gamma}}{\Gamma(\alpha)}
      +\frac{(-z)^{-\alpha}}{\Gamma(\gamma-\alpha)}\right) \qquad.
  \end{equation}
  Applying it into eq.\eqref{eq:point-lambda-wave-1} and
  eq.\eqref{eq:point-lambda-wave-2}, we get the asymptotic
  expansions of these wavefunctions when $a$ goes to infinity,
  \begin{eqnarray}
    \label{eq:point-lambda-wave-1-asymp-huge}
    \phi_1 &\sim &\Theta(\frac{3+\sqrt2}{6},-\nu) +\Theta(\frac{3+
    \sqrt2}{6},\nu) \qquad,\\
    \label{eq:point-lambda-wave-2-asymp-huge}
    \phi_2 &\sim &\Theta(\frac{3-\sqrt2}{6},-\nu) +\Theta(\frac{3-
    \sqrt2}{6},\nu) \qquad,
  \end{eqnarray}
  where
  \begin{equation}
    \Theta(\xi,\nu) = \Gamma(2\xi) \frac{e^{\nu
        a^3} a^{(-1/2+4\epsilon/\nu)}}{(2\nu)^{(\xi -4\epsilon/3\nu)}
      \Gamma(\xi+4\epsilon/3\nu)} \qquad.
  \end{equation}
  Since $\nu=2\sqrt{6\Lambda}/3$, the sign of $\Lambda$ will decide
  whether $\nu$ is real, and therefore decide how the asymptotic
  expansions behave.

  When $\Lambda<0$, $\nu=i(2\sqrt{-6\Lambda}/3)$ is an imaginary number
  and makes the modulus of both $\exp(\pm\nu a^3)$ and
  $a^{\pm 4\epsilon/\nu}$ being unit for any real $a$. So the
  asymptotic expansions can be simplified to
  \begin{equation}
    \phi_{1,2} \sim O(a^{-1/2}) \qquad.
  \end{equation}
  This descending with the order of minus one half is too slow, thus
  neither of the two eigen functions is normalizable.

  When $\Lambda>0$, with $\nu$ being a real number, the behavior of
  $\Theta$ is completely determined by the exponential term
  $\exp(\pm\nu a^3)$.  Fortunately, we know that $\Gamma(z)$ explodes
  at non-positive integer points, thus a carefully selected eigenvalue
  $\epsilon$ can make the exploded term $\Theta(\xi,\nu)$ in the
  expansions vanish.

  For $\phi_1$ it requires
  \begin{equation}
    \label{eq:point-lambda-wave-1-energy}
    \epsilon^{(1)}_n = -\frac{(6n+3+\sqrt2)}{2}\sqrt{\frac{\Lambda}{6}} \quad,\quad
    n=0,1,\dots \qquad.
  \end{equation}
  And we have the fact that for $F(\alpha,\gamma+1,z)$ whose $\alpha$
  is a non-positive integer, it can be described by Laguerre function
  \begin{equation}
    \label{eq:laguerre}
    L_n^{(\gamma)}(z) := {n+\gamma \choose n} F(-n,\gamma+1,z) \qquad.
  \end{equation}
  So the eigen states $|n^{(1)}\rangle$ for eigenvalue
  $\epsilon^{(1)}_n$ after normalization is written as
  \begin{equation}
    \label{eq:point-lambda-state-1}
    \langle a|n^{(1)}\rangle = c^{(1)}_n a^{1+\sqrt2/2} e^{-\nu a^3}
    L^{(\frac{\sqrt2}{3})}_n(2\nu a^3) \qquad,
  \end{equation}
  where $c^{(1)}_n$ is the normalization factor.

  Same discussion also holds for solution $\phi_2$, giving its
  eigenvalue
  \begin{equation}
    \label{eq:point-lambda-wave-2-energy}
    \epsilon^{(2)}_n = -\frac{(6n+3-\sqrt2)}{2}\sqrt{\frac{\Lambda}{6}} \quad,
    \quad
    n=0,1,\dots \qquad,
  \end{equation}
  and the normalized eigen state
  \begin{equation}
    \label{eq:point-lambda-state-2}
    \langle a|n^{(2)}\rangle = c^{(2)}_n a^{1-\sqrt2/2} e^{-\nu a^3}
    L^{(-\frac{\sqrt2}{3})}_n(2\nu a^3) \qquad,
  \end{equation}
  where $c^{(2)}_n$ is its normalization factor.

  From the definition of Kummer's function
  eq.\eqref{eq:kummer-function} we can see that for tiny arguments,
  \begin{eqnarray}
    \label{eq:point-lambda-wave-1-asymp-tiny}
    \phi_1(\epsilon\,;a) &\sim & a^{1+\sqrt2/2} (1-\nu a^3)(1
                                 +2\nu \frac{\alpha}{\gamma}a^3)
                                 \qquad,\\
    \label{eq:point-lambda-wave-2-asymp-tiny}
    \phi_2(\epsilon\,;a) &\sim & a^{1-\sqrt2/2} (1-\nu a^3)(1
                                     +\nu\frac{\alpha-\gamma+1}{2-
                                     \gamma}a^3) \qquad.
  \end{eqnarray}
  Therefore when $a$ goes to zero, $\phi_1$ and $\phi_2$ are all vanish,
  hence the cosmological singularity is naturally avoided.
  Thus we can choose the two sets of
  eigen states to be the basis of the quantum system:
  $\epsilon^{(1)}_n$, $|n^{(1)}\rangle$ and
  $\epsilon^{(2)}_n$, $|n^{(2)}\rangle$.

  For any real number $\alpha$, the first two Laguerre polynomials are
  \begin{eqnarray}
    \label{eq:laguerre-zero}
    L^{(\gamma)}_0 &= &1 \qquad, \\
    \label{eq:laguerre-one}
    L^{(\gamma)}_1 &= &1 +\gamma -x \qquad.
  \end{eqnarray}
  So the first set of eigen states with eigenvalue of the highest two are
  \begin{eqnarray}
    \label{eq:point-lambda-wave-2-highest}
    \langle a|0^{(1)}\rangle &= &c^{(1)}_0 a^{1+\sqrt2/2} e^{-\nu a^3}
    \qquad, \\
    \label{eq:point-lambda-wave-1-highest}
    \langle a|1^{(1)}\rangle &= &c^{(1)}_1 a^{1+\sqrt2/2} \left(\frac{3+
    \sqrt2}
    {3}-2\nu a^3\right)
                           e^{-\nu a^3} \qquad.
  \end{eqnarray}
  If a certain state is a combination of only these two states
  $|\psi\rangle=\alpha|0^{(1)}\rangle+\beta|1^{(1)}\rangle$ at
  initial~($|\alpha|^2+|\beta|^2=1$), it will evolve with respect to
  the cosmological time $t$ as
  \begin{equation}
    \label{eq:point-lambda-wave-highest-comb}
    |\psi,t\rangle = \alpha|0^{(1)}\rangle e^{-i\epsilon^{(1)}_0 t}
    +\beta|1^{(1)}\rangle e^{-i\epsilon^{(1)}_1 t} \qquad.
  \end{equation}
  The evolution of the average measurement $\overline{a^2}$ can be
  calculated out:
  \begin{align*}
    \overline{a^2}(t)= & \langle\psi,t|a^2|\psi,t\rangle \\
    = & |\alpha|^2\langle0|a^2|0\rangle
    +\alpha^*\beta \langle0|a^2|1\rangle e^{i(\epsilon^{(1)}_0-
    \epsilon^{(1)}_1)t}
    +|\beta|^2\langle1|a^2|1\rangle
    +\alpha\beta^* \langle1|a^2|0\rangle e^{-i(\epsilon^{(1)}_0-
    \epsilon^{(1)}_1)t} \\
    = & 2 \Re(\alpha^*\beta e^{i(\epsilon^{(1)}_0-\epsilon^{(1)}_1)t})
    \langle0|a^2|1\rangle
    +(|\alpha|^2\langle0|a^2|0\rangle +|\beta|^2\langle1|a^2|1\rangle) \\
    = &2c_{0,1}|\alpha||\beta|
    \cos\left(\frac{\sqrt{6\Lambda}}{2}\,t -\theta\right)
    +(c_{0,0}|\alpha|^2 +c_{1,1}|\beta|^2) &\qquad.
  \end{align*}
  where $c_{n,m}=\langle n^{(1)}|a^2|m^{(1)}\rangle$,
  $\theta=\operatorname{Arg}(\alpha)-\operatorname{Arg}(\beta)$.
  Fortunately, a certain state which is a combination of $|n^{(2)}\rangle$ can
  get the similar result as $(\epsilon_{n}^{(2)}-\epsilon_{m}^{(2)})$ is also
  always some integer times of $\sqrt{6\Lambda}/2$ according to
  eq.\eqref{eq:point-lambda-wave-2-energy}.

  Therefore it is reasonable to assuming that our universe is in a state of
  combination of $|n^{(1)}\rangle$ or $|n^{(2)}\rangle$. This solution
  suggests a pulsing universe with a characteristic time
  of $4\pi\sqrt{1/6\Lambda}$.  Actually although we do not know which
  state the universe is at a certain time, we can prove that it always
  rebounds with the same characteristic time no matter how the initial
  coefficients for the state of each energy level are given.  It is
  clear that the evolution of $\overline{a^2}$ is always described by
  \begin{equation}
    \label{eq:point-lambda-average-a-normal}
    \overline{a^2}(t) = \sum_{n,m\geqslant0} \eta_{nm}
    \cos((\varepsilon_n -\varepsilon_m)t +\theta_{nm})
    \qquad.
  \end{equation}
  We can see
  $\varepsilon_n-\varepsilon_m$ is always some integer times of
  $\sqrt{6\Lambda}/2$ which ensures that $\overline{a}(t)$ has a period
  of $4\pi\sqrt{1/6\Lambda}$.

  Considering the universe as we observed is experiencing an
  accelerating expansion now, it is
  reasonable to assume it is still in the first quarter of the period,
  implying the cosmological constant $\Lambda$ should be no bigger
  than $\pi^2/6T_0^2$ where $T_0$ stands for the cosmological time
  till now.

  Another interesting thing is that, observing the combination of
  solutions can provide a square wave, this quantum system must have
  some special states that may let $\overline{a^2}$ rise as fast as
  possible at some certain time $t_0$. For the simplest case,
  considering a state composed by eigen states with real coefficients
  $|\psi,t\rangle=\sum_n\tau_ne^{-i\varepsilon_n t}|n\rangle$,
  if its scale factor evolves like
  \begin{equation}
    \label{eq:square}
    \overline{a^2}(t) = A \sum_{k=0}^{N} \frac{1}{2k-1}
    \cos\left(\frac{\sqrt{6\Lambda}}{2}(2k-1)\,t\right) +C \qquad,
  \end{equation}
  then choose only the coefficients of the first $2N-1$ states to be
  non-zero and real, we know they will satisfy the polynomial system
  \begin{equation}
    \label{eq:polynomial-coefficient}
    \sum_{n=0}^{2N-l-1} \tau_n\tau_{n+l}c_{n,n+l} = \frac{A}{4}
    \left(\frac{1-(-1)^l}{l}\right) \quad,\quad 1\leqslant l\leqslant
    2N-1 \qquad,
  \end{equation}

  There are in total $2N$ coefficients needed to be fixed. The
  normalization condition together with
  eq.\eqref{eq:polynomial-coefficient} exactly give the same number of
  equations from which the coefficients can be solved. Therefore we
  can satisfy eq.\eqref{eq:square} for any $N$ as large as we
  wish. That gives the possibility of an expansion of $a$ at any
  velocity, which may generate an inflation with the speed even faster
  than exponential level as normal understanding.
  
  \section{Quantum Cosmology of $f(R)$ Gravity}
  For more general cases, we consider the quantum model of a flat FRW
  universe under $f(R)$ gravity. More detailed discussion on the
  point-like model of $f(R)$ universe can be found in the works of
  Capozziello~\cite{capozziello_fr_2008}. The point-like Lagrangian
  with no term of matter will be like
  \begin{equation}
    \label{eq:point-fR-lagrangian}
    L = (f-f_RR)a^3 -6f_{RR}\dot{R}a^2\dot{a}
    -6f_Ra\dot{a}^2 \qquad,
  \end{equation}
  and the canonical momentums for $a$ and $R$ are respectively
  \begin{eqnarray}
    \label{eq:point-fR-momentum-a}
    \pi_a &= &-6f_{RR}\dot{R}a^2 -12f_Ra\dot{a} \qquad,\\
    \label{eq:point-fR-momentum-R}
    \pi_R &= &-6f_{RR}a^2\dot{a} \qquad.
  \end{eqnarray}
  So the canonical energy is
  \begin{equation}
    \label{eq:point-fR-hamiltonian}
    E_{L} = -(f-f_RR)a^3 -6f_{RR}\dot{R}a^2\dot{a}^2
    -6f_Ra\dot{a}^2 \qquad.
  \end{equation}

  From eq.\eqref{eq:point-fR-momentum-R} we can directly read that
  \begin{equation}
    \dot{a} = -\frac{\pi_R}{6f_{RR}a^2} \qquad,
  \end{equation}
  Plug it into eq.\eqref{eq:point-fR-momentum-a} and get
  \begin{equation}
    6f_{RR}\dot{R}a^2 = -\pi_a +\frac{2}{a}\frac{f_R}{f_{RR}}\pi_R \qquad.
  \end{equation}
  Applying them to eq.\eqref{eq:point-fR-hamiltonian}, the Hamiltonian
  for the system becomes
  \begin{align*}
    H &= -(f-f_RR)a^3 +\frac{\pi_R}{6f_{RR}a^2} \left(-\pi_a
                  +\frac{2}{a}\frac{f_R}{f_{RR}}\pi_R\right) -6f_Ra
                  \left(-\frac{\pi_R}{6f_{RR}a^2}\right)^2 \\
                &= -(f-f_RR)a^3
                  -\frac{1}{6a^2}\frac{1}{f_{RR}}\pi_R\pi_a
                  +\frac{1}{6a^3}\frac{f_R}{f_{RR}^2}\pi_R^2 \qquad.
  \end{align*}
  So the wave equation that describes this quantum system
  is
  \begin{align*}
    \label{eq:point-fR-schrodinger}
  i\frac{\partial}{\partial t}\Psi = -\frac{1}{6a^3}\frac{f_R}{f_{RR}^2}\frac{\partial^2}{\partial R^2}\Psi+\frac{1}{6a^2}\frac{1}{f_{RR}}\frac{\partial^2}{\partial a \partial R}\Psi+\frac{1}{24a^3}\left(\frac{6f_Rf_{RRR}}{f_{RR}^3}-\frac{7}{f_{RR}}\right)\frac{\partial}{\partial R}\Psi-
  \frac{1}{12a^2}\\
  \frac{f_{RRR}}{f_{RR}^2}\frac{\partial}{\partial a}\Psi
  -\left[\frac{1}{24a^3}\left(\frac{6f_Rf_{RRR}^2}{f_{RR}^4}-\frac{5f_{RRR}}{f_{RR}^2}-\frac{2f_Rf_{RRRR}}{f_{RR}^3}\right)+a^3(f-f_RR)\right]\Psi \qquad.
  \end{align*}
  $\Psi(t,a,R)$ is a function of cosmological time $t$, scale factor
  $a$ and Ricci scalar $R$.

  The equation relies on the form of $f(R)$ to be exactly solved.
  However as a linear partial differential equation, its coefficients
  of all the second order terms satisfy the fact that
  $\Delta=1/(144a^4f_{RR}^2)$ is positive on the whole $a-R$ plane.
  Therefore the eigen equation of the operator $\hat{H}$ is a
  hyperbolic equation and can be transformed into a wave equation.

  We need to point out that $a$ and $R$ have been separated via
  Palatini formalism. In this case, their relation is linked by one of
  the equations of motion rather than a given definition. So after the
  quantization, this relation must have degenerated to be
  statistically satisfied.  That means even a flat universe of small
  scale or a huge scale universe with large curvature which are not
  normally allowed in the classical case will also have contribution
  to the possibility.
  
  \section{Conclusion And Discussions}
  The purpose of this paper is to introduce a new approach to
  inquire the minisuperspace model without seeking the Wheeler-DeWitt
  equation for a certain gravitational theory.

  The technique is to apply the cosmological principle directly to the
  action of a gravitational system before variation, and reform the
  Lagrangian of geometry to a classical point-like one. It is obvious
  that such a process of taking the metric of a cosmological model
  which is truncated by an enormous degree of imposed symmetry and
  simply plugging it into a quantization procedure should not give an
  answer that is in any way an exact solution. However, strange
  enough, we have seen that the variation of this point-like
  Lagrangian gives the right equation of motion~(the Friedmann
  equation) to describe the universe.
  
  By quantizing this semi-classical system described by the point-like
  Lagrangian, we represent a quantum system that is very similar to
  the minisuperspace from reducing the superspace where the
  Wheel-DeWitt equation is defined on. The only difference is, in our
  situation, for solving a semi-classical Schr\"{o}dinger equation we
  need the concept of the eigenvalue $\epsilon$ of the Hamiltonian of
  the system which does not exist in the classical minisuperspace
  theory.
  
  It is very natural to apply our technique beyond the Einstein
  gravity to the gravitational model with cosmological constant and
  more general $f(R)$ gravity with the help of Palatini formalism and
  respectively get their quantum cosmological model. This especially
  opens the gate for considering quantum cosmology of $f(R)$ gravity.

  As the second aspect of our work shown in this paper, we give the
  exact solutions of the quantum systems we get under Einstein gravity
  with and without cosmological constant. We find that the existence 
  of a tiny positive cosmological constant is reasonable

  We prove that all possible states in such a legal quantum
  cosmological model predict pulsing universe with the same period of
  a cosmological characteristic time that is inversely proportional to
  the square root of the cosmological constant. Considering the
  enormous amount of time the universe has existed, the cosmological
  constant must be extremely tiny.

  Moreover, we show that this quantum system contains states that allow
  expansion at any speed as fast as possible, which could probably
  provide a motivation for inflation.
  
  \section{Acknowledgments}

  We would like to thank the National Natural Science Foundation of
  China~(Grant No.11571342) for supporting us on this work.
%\begin{figure}
%\resizebox{0.75\textwidth}{!}{%
%  \includegraphics{leer.eps}
%\vspace{5cm}       % Give the correct figure height in cm
%\end{figure}
%\begin{figure*}
%\vspace*{5cm}       % Give the correct figure height in cm
%\end{figure*}
%\begin{table}
%\begin{tabular}{lll}
%\hline\noalign{\smallskip}
%\end{tabular}
% Or use
%\vspace*{5cm}  % with the correct table height
%\end{table}
%
% BibTeX users please use
 \bibliographystyle{unsrt}
 \bibliography{reference}

\begin{thebibliography}{10}

\bibitem{dewitt_quantum_1967}
Bryce~S. DeWitt.
\newblock Quantum {Theory} of {Gravity}. {I}. {The} {Canonical} {Theory}.
\newblock {\em Physical Review}, 160(5):1113--1148, August 1967.

\bibitem{misner_quantum_1969}
Charles~W. Misner.
\newblock Quantum {Cosmology}. {I}.
\newblock {\em Physical Review}, 186(5):1319--1327, October 1969.

\bibitem{hawking_boundary_1982}
S.~W. Hawking.
\newblock The {Boundary} {Conditions} of the {Universe}.
\newblock {\em Pontif.Acad.Sci.Scr.Varia}, 48:563--574, 1982.

\bibitem{hartle_wave_1983}
J.~B. Hartle and S.~W. Hawking.
\newblock Wave function of the {Universe}.
\newblock {\em Physical Review D}, 28(12):2960--2975, December 1983.

\bibitem{halliwell1}
Jonathan~J. Halliwell and Jorma Louko.
\newblock Steepest-descent contours in the path-integral approach to quantum
  cosmology. i. the de sitter minisuperspace model.
\newblock {\em Phys. Rev. D}, 39:2206--2215, Apr 1989.

\bibitem{halliwell2}
Jonathan~J. Halliwell and Jorma Louko.
\newblock Steepest-descent contours in the path-integral approach to quantum
  cosmology. ii. microsuperspace.
\newblock {\em Phys. Rev. D}, 40:1868--1875, Sep 1989.

\bibitem{halliwell3}
Jonathan~J. Halliwell and Jorma Louko.
\newblock Steepest-descent contours in the path-integral approach to quantum
  cosmology. iii. a general method with applications to anisotropic
  minisuperspace models.
\newblock {\em Phys. Rev. D}, 42:3997--4031, Dec 1990.

\bibitem{PhysRevD.42.1091}
R.~de~Ritis, G.~Marmo, G.~Platania, C.~Rubano, P.~Scudellaro, and
  C.~Stornaiolo.
\newblock New approach to find exact solutions for cosmological models with a
  scalar field.
\newblock {\em Phys. Rev. D}, 42:1091--1097, Aug 1990.

\bibitem{PhysRevD.69.103510}
Claudio Rubano, Paolo Scudellaro, Ester Piedipalumbo, Salvatore Capozziello,
  and Monica Capone.
\newblock Exponential potentials for tracker fields.
\newblock {\em Phys. Rev. D}, 69:103510, May 2004.

\bibitem{PhysRevD.83.103512}
Spyros Basilakos, Michael Tsamparlis, and Andronikos Paliathanasis.
\newblock Using the noether symmetry approach to probe the nature of dark
  energy.
\newblock {\em Phys. Rev. D}, 83:103512, May 2011.

\bibitem{PhysRevD.62.043506}
Ruggiero de~Ritis, Alma~A. Marino, Claudio Rubano, and Paolo Scudellaro.
\newblock Tracker fields from nonminimally coupled theory.
\newblock {\em Phys. Rev. D}, 62:043506, Jul 2000.

\bibitem{sanyal_general_2003}
A.K. Sanyal, C.~Rubano, and E.~Piedipalumbo.
\newblock Coupling parameters and the form of the potential via noether
  symmetry.
\newblock {\em General Relativity and Gravitation}, 35(9):1617--1635, 2003.

\bibitem{PhysRevD.52.3288}
S.~Capozziello, M.~Demianski, R.~de~Ritis, and C.~Rubano.
\newblock Cosmological perturbations in exact-noether background solutions.
\newblock {\em Phys. Rev. D}, 52:3288--3297, Sep 1995.

\bibitem{10.4137/GBI.S4273}
Hong Lu, Lynda~M. McDowell, Daniel~R. Studelska, and Lijuan Zhang.
\newblock Glycosaminoglycans in human and bovine serum: Detection of
  twenty-four heparan sulfate and chondroitin sulfate motifs including a novel
  sialic acid-modified chondroitin sulfate linkage hexasaccharide.
\newblock {\em Glycobiology Insights}, 2:13--28, 02 2010.

\bibitem{0264-9381-27-13-135019}
Yi~Zhang, Yun gui Gong, and Zong-Hong Zhu.
\newblock The noether symmetry approach in a 'cosmic triad' vector field
  scenario.
\newblock {\em Classical and Quantum Gravity}, 27(13):135019, 2010.

\bibitem{0264-9381-25-22-225006}
Rudinei~C de~Souza and Gilberto~M Kremer.
\newblock Noether symmetry for non-minimally coupled fermion fields.
\newblock {\em Classical and Quantum Gravity}, 25(22):225006, 2008.

\bibitem{wei_noether_2012}
Hao Wei, Xiao-Jiao Guo, and Long-Fei Wang.
\newblock Noether {Symmetry} in \$f({T})\$ {Theory}.
\newblock {\em Physics Letters B}, 707(2):298--304, January 2012.
\newblock arXiv: 1112.2270.

\bibitem{capozziello_fr_2008}
Salvatore Capozziello and Antonio De~Felice.
\newblock f({R}) cosmology by {Noether}'s symmetry.
\newblock {\em Journal of Cosmology and Astroparticle Physics}, 2008(08):016,
  August 2008.

\bibitem{capozziello_general_2000}
S.~Capozziello and G.~Lambiase.
\newblock Higher-order corrections to the effective gravitational action from
  noether symmetry approach.
\newblock {\em General Relativity and Gravitation}, 32(2):295--311, 2000.

\bibitem{sanyal_general_2011}
AbhikKumar Sanyal, Claudio Rubano, and Ester Piedipalumbo.
\newblock Noether symmetry for gauss–bonnet dilatonic gravity.
\newblock {\em General Relativity and Gravitation}, 43(10):2807--2820, 2011.

\bibitem{weyl1950theory}
Hermann Weyl.
\newblock {\em The theory of groups and quantum mechanics}.
\newblock Courier Corporation, 1950.

\end{thebibliography}

%\begin{thebibliography}{}
%Author, \textit{Book title} (Publisher, place year) page numbers
% etc
%\end{thebibliography}
\end{document}